\documentstyle[aps, prb]{revtex}
\include{epsf}

\newcommand{\tlbacod}{Tl$_2$Ba$_2$CuO$_{6+\delta}$}
\newcommand{\itM}{\mbox{\boldmath $M$}}
\newcommand{\itH}{\mbox{\boldmath $H$}}
\newcommand{\ittau}{\mbox{\boldmath $\tau$}}
\newcommand{\degrees}[1]{\mbox{$#1{}^{\circ}$}}

\begin{document}
\draft

\twocolumn[\hsize\textwidth\columnwidth\hsize\csname@twocolumnfalse\endcsname

\title{Superconducting Magnetization above the Irreversibility Line in
  \tlbacod}

\author{C.~Bergemann}
\address{Cavendish Laboratory, University of Cambridge,
  Madingley Road, Cambridge~CB3~0HE, United~Kingdom}

\author{A.~W.~Tyler,\footnotemark[1] A.~P.~Mackenzie,\footnotemark[1]
  and J.~R.~Cooper} 
\address{Interdisciplinary Research Centre in
  Superconductivity, University of Cambridge, Madingley Road,
  Cambridge~CB3~0HE, United~Kingdom}

\author{S.~R.~Julian}
\address{Cavendish Laboratory, University of Cambridge,
  Madingley Road, Cambridge~CB3~0HE, United~Kingdom}

\author{D.~E.~Farrell\footnotemark[2]}
\address{Interdisciplinary Research Centre in Superconductivity,
  University of Cambridge, Madingley Road, Cambridge~CB3~0HE,
  United~Kingdom}
        
\date{Revised manuscript, submitted to Physical Review B on 21~January~1998}
\maketitle
\begin{abstract}
  Piezolever torque magnetometry has been used to measure the
  magnetization of superconducting \tlbacod. Three crystals with
  different levels of oxygen overdoping were investigated in magnetic
  fields up to 10~Tesla. In all cases, the magnetization above the
  irreversibility line was found to depart from the behaviour \mbox{$M
    \sim \ln (\eta H_{c2} / H)$} of a simple London-like vortex
  liquid. In particular, for a strongly overdoped \mbox{($T_c = 15$K)}
  crystal, the remnant superconducting order above the irreversibility
  line is characterized by a linear diamagnetic response \mbox{($M
    \sim H$)} that persists well above $T_c$ and also up to the
  highest field employed.
\end{abstract}
\pacs{PACS numbers: 74.25.Ha, 74.60.Ec, 74.72.Fq, 07.55.Jg}
\vspace{.8cm}]

\section{Introduction}

A distinguishing feature of the high-$T_c$ cuprate superconductors is
that a finite resistance (in the zero current limit) appears at a
magnetic field well below that required to restore the full normal
state resistance.  Experimentally, this resistance onset field is
found to coincide with the so-called irreversibility field $H_{\rm
  irr}$, above which magnetic irreversibility vanishes. A broad
superconducting to normal transition is thought to reflect the
existence of a mobile vortex liquid between $H_{\rm irr}$ and
$H_{c2}$.\cite{ginsberg} Although diffraction evidence for a genuine
vortex liquid has yet to be reported, the term will be used here to
designate the state of the vortex assembly between these two
characteristic fields. Physically, the existence of a liquid is
thought to be a consequence of large anisotropy and a short
superconducting coherence length, both factors weakening the
vortex-vortex interaction. In one of the many available
scenarios,\cite{ginsberg} this weak flux lattice is thought to be
``melted'' at $H_{\rm irr}$ by strong thermal fluctuations.

An open question is how this behaviour evolves as $T_c$ is lowered and
the coherence length increases. \tlbacod\ is an attractive compound
for an investigation of this issue: with an optimal critical
temperature of around 90K, it exists in a comparatively simple single
layer structure, and strongly overdoped tetragonal single crystals can
be produced by introducing relatively little excess oxygen \mbox{($T_c
  \to 0 \mbox{ for } \delta = 0.1$)}.\cite{kubo,manako,liu} In some
respects, the overdoping that is possible in \tlbacod\ is unique: for
example, overdoping the La$_{2-x}$Sr$_x$CuO$_4$ system leads to high
levels of disorder. In \tlbacod, the disorder associated with doping
resides in or between the TlO~bilayers, so disorder-related carrier
scattering in the CuO$_2$ planes is relatively weak; in-plane
resistivities just above $T_c$ can be less than 10\,$\mu\Omega$cm,
even in strongly overdoped material.\cite{hstar} A variety of
experiments have been performed on tetragonal single crystals and
epitaxial thin films. For example, the superconducting gap symmetry
has been assessed in tri- and quadri-crystal thin
films\cite{tsuei1,tsuei2} and in microwave experiments,\cite{broun}
the low-field vortex properties of the vortex lattice have been
studied by magnetization\cite{zuo} and Bitter patterning,\cite{pardo}
and a number of normal state properties have also been
investigated.\cite{manako,magnetores}

In \tlbacod\ crystals with high values of $T_c$, resistance
measurements\cite{carrington1} have shown that the
normal-superconducting transition is broadened, as observed in other
cuprate materials.  However, for strongly overdoped material with low
$T_c$ values, sharp resistive transitions are observed.\cite{hstar}
These are similar in appearance to the transitions encountered in
conventional superconductors, where it is known that \mbox{$H_{\rm
    irr} \sim H_{c2}$}.\cite{ginsberg} This suggests that lowering
$T_c$ by overdoping somehow eliminates the flux liquid from the vortex
phase diagram.

On the other hand, the temperature dependence of the apparent upper
critical field deduced from the overdoped resistivity data is
strikingly different from that expected in a conventional
superconductor: it exhibits an upward curvature from \mbox{$T_c =15$K}
down to the lowest temperatures studied (12mK) and shows no sign of
saturation for \mbox{$T \to 0$}.\cite{hstar} Qualitatively similar
observations have been reported on thin films of
Bi$_2$Sr$_2$CuO$_6$\cite{osofsky} and other hole doped cuprates. This
anomalous apparent $H_{c2}$ in the hole doped materials has attracted
substantial attention. Theoretical proposals fall into two broad
classes. One of these\cite{kotliar,cooper} regards the existence of
residual superconducting order above the apparent $H_{c2}$ as an
essential aspect of the phenomenon, while the other
approach\cite{alexandrov,rasolt,dias,ovchinnikov,koyama,schofield,abrikosov}
considers the transition seen in the resistivity as an accurate
estimate of the location of the transition from the superconducting to
the normal state.

The sharpness of the transition and the absence of a measurable
magnetoresistance at higher fields both argue against the existence of
residual superconducting order. However, it is difficult to estimate
quantitatively the contribution of such order to the conductivity, so
transport measurements by themselves cannot decide this issue.
Studies of the specific heat or magnetization would be more
definitive, but single crystals of high quality tend to be small
\mbox{(${\rm mass} < 10^{-5}$grams)} and the experiments
correspondingly difficult. Nonetheless, Carrington et
al.\cite{carrington2} have recently succeeded in measuring the
specific heat of a small single crystal with \mbox{$T_c = 17$K}.
Although an applied field dramatically reduces the size of the
superconducting specific heat anomaly, the results imply that remnant
superconducting order persists well above the apparent~$H_{c2}$.

In summary, the information available on the vortex liquid in
\tlbacod\ presents a seemingly conflicting picture. As the transition
temperature is reduced by overdoping, specific heat measurements
suggest that a wide vortex liquid region persists. By contrast, the
sharp resistive transitions suggest that the liquid region becomes
very narrow in highly overdoped material. Magnetization offers the
most direct probe of vortex behaviour, but previous SQUID efforts to
measure this quantity on highly overdoped material have been hampered
by inadequate sensitivity.\cite{andymacunpub}

\begin{figure}
\centerline{\epsfxsize=8cm\epsfbox{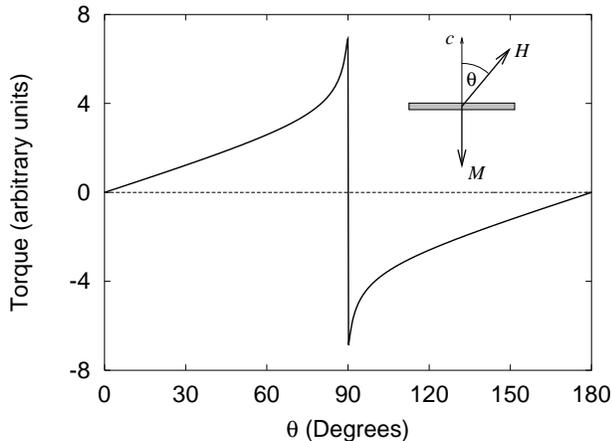}}
\caption{Predicted angular dependence of the torque (in arbitary
  units) according to Eq.~(\ref{kogantorque}), with \mbox{$H/\eta
    H_{c2}$ = 0.1}. Orientations of the applied field \itH, the
  magnetization \itM, and the $c$-axis are shown in the inset.}
\label{koganplot}
\end{figure}

We have employed a recently introduced variant of torque magnetometry
-- using piezoresistive microcantilevers -- to reach a working
resolution of $10^{-13}{\rm Am}^2$ in a field of 10~Tesla. This is
about two orders of magnitude more sensitive than is available from
standard SQUID techniques, allowing us to explore the magnetization of
\tlbacod\ over a wide doping range. We find that a diamagnetic torque
persists above $H_{\rm irr}$ for all the crystals examined. However,
the underlying magnetization departs drastically from that of a London
liquid.

\section{Torque Principles} \label{TorqPrinc}

For a material in which the magnetization \itM\ is uniform, the
magnetic torque density (torque per unit volume) is given by
\mbox{$\ittau = \itM\times\itH$}, where \itH\ is the applied magnetic
field. If \itH\ is applied along a symmetry axis of a single crystal,
the equilibrium magnetization is parallel to \itH, and the torque is
zero. The magnetic response of tetragonal high-$T_c$ materials to
off-axis fields is largely controlled by their superconducting
anisotropy, \mbox{$\gamma = (m_c/m_a)^{1/2}$}. Here, $m_c$ and $m_a$
are the Ginzburg-Landau superconducting effective masses for pair
motion along the $c$-direction, and in the CuO$_2$~planes,
respectively. Both the magnitude and direction of \itM\ depend on
$\gamma$ and on the angle $\theta$ that the field makes with the
$c$-axis. The anisotropy of all the crystals studied in this work was
large \mbox{($\gamma > 50$)}, simplifying the interpretation of the 
data.

The torque on an assembly of vortices in a high-$T_c$ superconductor
exhibits a characteristic angular dependence.\cite{kogan1} Its origin
may be understood by referring to Fig.~\ref{koganplot}. For
large~$\gamma$, the magnetization lies very close to the $c$-axis, as
shown in the inset, while its magnitude $M$ depends only on the
component of the applied field along that axis. The supposition that
$\itM$ is fully determined by the effective field $H\cos\theta$ is a
fundamental result of scaling analysis\cite{blatter} in the
large-$\gamma$ limit of the anisotropic Ginzburg-Landau model.

Under these circumstances, and for \mbox{$H_{c1} \ll H \,
  |\!\cos\theta| \ll H_{c2}$}, the equilibrium vortex magnetization
can usually be well approximated by the London result \mbox{$M =
  (\Phi_0/8\pi\mu_0\lambda^2) \: \ln\left(\eta
    H_{c2}/H\,|\!\cos\theta|\right)$}, where $\lambda$ and $H_{c2}$
are the in-plane penetration depth and the upper critical field along
the $c$-axis, respectively, and $\eta$ is a numerical parameter of
order unity.\cite{kogan1} For \mbox{$0 < \theta < \pi/2$}, the
magnetic moment lies along the {\em negative\/} $c$-axis, and the
magnitude of the vortex torque density is given by
\begin{equation} \label{kogantorque}
  \tau_{\rm v}(\theta) = MH \sin\theta = \frac{\Phi_0
                    H\sin\theta}{8\pi\mu_0\lambda^2} \: \ln\left(
                    \frac{\eta H_{c2}}{H \, |\!\cos\theta|} \right)
                    \,.
\end{equation}
For \mbox{$\theta > \pi/2$}, the magnetization points along the {\em
  positive\/} $c$-axis, and the sign of the torque is reversed. At
\mbox{$\theta = \pi/2$}, the magnetic field crosses the CuO$_2$
planes, and the superconducting screening currents abruptly change
direction.

A more complete description of the vortex torque\cite{kogan1}
indicates that Eq.~(\ref{kogantorque}) is correct to \mbox{$\pm 2\%$}
for \mbox{$\gamma = 50$} and for field orientations lying more than
\degrees{5} from the CuO$_2$ planes. A typical predicted torque
characteristic is shown in Fig.~\ref{koganplot}. The drastic increase
of $\tau$ at high angles is due to the logarithmic term in
Eq.~(\ref{kogantorque}): as the effective applied field decreases, the
diamagnetic magnetization of the crystal increases, reaching a sharp
peak as the crystal enters the Meissner state at $H_{c1}$.

In traditional magnetization measurements, the irreversibility field
is probed by cycling the magnetic field up and down and recording the
value at which irreversibility develops. Since the irreversibility is
anisotropic, the symbol $H_{\rm irr}$ is reserved here to denote the
$c$-axis irreversibility field. For reasons discussed in the Appendix,
$H_{\rm irr}$ was measured here in a different manner: the field was
fixed at some value $H$, while the angle~$\theta$, and with it the
effective field $H\cos\theta$, was cycled. Angular hysteresis is
observed above some angle $\theta_{\rm irr}$.  Assuming a large
anisotropy, $H_{\rm irr}$ can then be deduced from directly measured
quantities, viz.
\begin{equation} \label{thetairr}
  H_{\rm irr} = H \cos\theta_{\rm irr} \:.
\end{equation}

\section{Experimental}

The critical temperatures of the three \tlbacod\ crystals used in this
work were established as the temperatures above which the low-field
superconducting vortex torque signature disappeared. The values
obtained, \mbox{$T_c = 15$K}, 25K, and 85K, agreed with previous
estimates from resistivity measurements for the \mbox{$T_c = 15$K}
sample\cite{hstar} and with the specific heat anomalies observed in
the other two crystals\cite{marcenat} within experimental uncertainty
(dominated by the superconducting transition widths) of a few percent.

All three samples were grown using a self-flux method in alumina
crucibles.\cite{tyler} As-grown crystals are overdoped, so the desired
doping level was established by low temperature \mbox{($T <
  \degrees{400}$C)} annealing in various atmospheres.\cite{tyler} For
our torque experiments, the samples were then cut into platelets with
typical dimensions \mbox{$160 \times 80 \times 10 \, \mu {\rm m}^3$}.

Torque measurements were performed using piezoresistive
microcantilevers. Originally designed for atomic force
microscopy,\cite{tortonese} their use as sensitive torque sensors was
pioneered by Rossel and coworkers.\cite{rossel} Commercial silicon
piezolevers\cite{psi} were employed in this work, their dimensions
being \mbox{$170 \times 50 \times 5 \, \mu {\rm m}^3$}. A boron-doped
path implanted on the upper lever surface has a resistance of about
3k$\Omega$ at 4K\@. On applying a torque, the resistance changes by a
small amount \mbox{$\Delta R$}. A typical number for the response is
\mbox{$\Delta R/R \sim 10^7$} per~Nm.

Using an $XYZ$ micropositioner, the crystal was epoxied to the end of
a cantilever, its CuO$_2$ planes coinciding with the flat surface of
the lever. A second (empty) piezolever was employed to compensate
background signals, using a Wheatstone bridge circuit driven by a
floating 77~Hz AC~current source. The current (50$\mu$A through each
lever) increased the sample temperature by less than 0.1K, at all
temperatures.

The two levers were mounted closely together on a rotation stage
inside a pumped $^4$He cryomagnetic system capable of reaching 1.3K in
a 15T field. The magnet was operated in persistent mode, and its field
inside the rotation stage was homogeneous to \mbox{$\sim 0.1\%$}. The
cantilevers were either in $^4$He exchange gas or immersed directly in
liquid helium.

Both calibration and uncertainty estimation are more involved than for
traditional torque methods.\cite{beck} The main source of random
uncertainty comes from variations in temperature. A systematic
background signal originates from the varying gravitational torque on
the crystal when the sample stage is rotated. However, the largest
measurement uncertainty stems from the intrinsic magnetoresistance of
the levers. Over most of our measurement range, the torque density
data presented in the next section are estimated to be reliable to
\mbox{$\pm 10\,$N/m$^2$}. For a more detailed discussion of the
measurement uncertainties, the reader is referred to
Appendix~\ref{errest}.

\section{Results}

\subsection{Anisotropy and Irreversibility}

In principle, the superconducting anisotropy $\gamma$ may be
determined from the angular dependence of the torque when the field
lies close to the CuO$_2$ planes. In practice, the finite mosaic
spread of the crystal broadens the high-angle characteristic. For
this, and other reasons, the data can only be used to establish a
lower bound for $\gamma$. These difficulties have been discussed in
the literature.\cite{beck} Nonetheless, for all three crystals, we
estimate that \mbox{$\gamma > 50$}, justifying the high-anisotropy
approximation described in section~\ref{TorqPrinc}.

\begin{figure}
\centerline{\epsfxsize=8cm\epsfbox{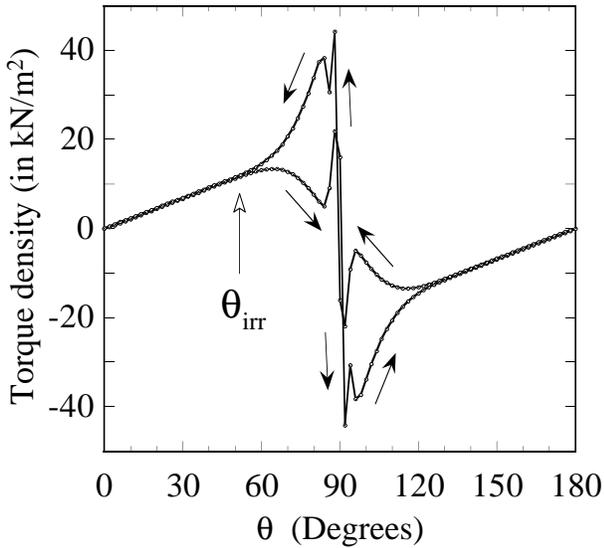}}
\caption{Angular dependence of the torque density for the \mbox{$T_c =
    85$K} crystal in an applied field of 1~Tesla and at a temperature
  of 35.3K\@. The direction of angular sweep is indicated by the arrows.
  Irreversibility occurs above the angle identified as $\theta_{\rm
    irr}$. The complex behaviour inside the irreversible region will
  not be discussed here.}
\label{irrplot}
\end{figure}

\begin{figure}
\centerline{\epsfxsize=7.98cm\epsfbox{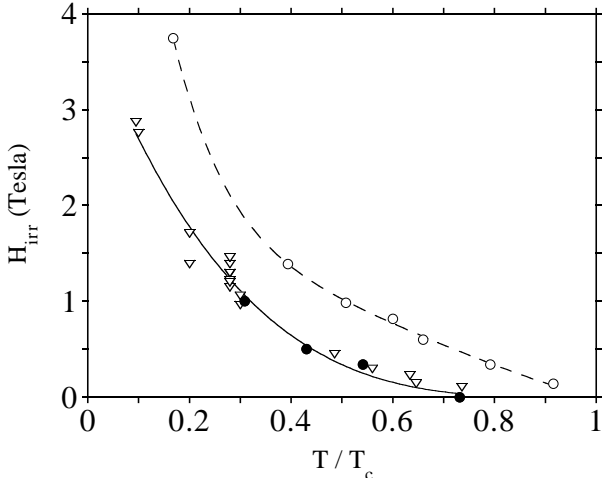}}
\caption{The irreversibility field along the $c$-axis, $H_{\rm irr}$,
  as a function of reduced temperature, $T/T_c$, for the \mbox{$T_c =
    15$K} (open triangles), 25K (open circles), and 85K crystal (full 
  circles).}
\label{irrfields}
\end{figure}

Fig.~\ref{irrplot} shows the angular dependence of the torque density
observed for the \mbox{$T_c = 85$K} crystal at \mbox{$T = 35.3$K} in a
field of 1~Tesla. (Note that in this and subsequent figures showing
angular dependences, the data have been taken over two angular
quadrants and have then been symmetrized according to
Eq.~(\ref{symmetrize}), as discussed in the Appendix). The
irreversibility angle $\theta_{\rm irr}$, identified on this Figure,
allows the associated irreversibility field $H_{\rm irr}$ to be
estimated using Eq.~(\ref{thetairr}). $H_{\rm irr}(T)$ was obtained in
this manner for all three crystals and is shown in
Fig.~\ref{irrfields}. The data for the \mbox{$T_c = 15$K} crystal are
in good agreement with previous reports of the ``foot'' of the
resistive transition in a magnetic field.\cite{hstar} Also, the
irreversibility fields obtained with the angular sweep technique
agreed with control runs at selected temperatures that used a simple
field sweep with $\itH$ pointing close to (but not along) the
$c$-axis.

Note that the irreversibility lines for the \mbox{$T_c = 15$K} and the
\mbox{$T_c = 85$K} samples almost coincide when plotted against
reduced temperature. We believe that this is an intrinsic property of
the material -- it should again be stressed that the samples used in
our study are of high quality, as indicated by their sharp specific
heat anomalies.\cite{marcenat} Also, the lack of an unambiguous trend
of the irreversibility fields with~$T_c$ is again in accordance with
resistivity measurements across large parts of the doping
range.\cite{carrington1} Still, this behaviour is quite surprising and
warrants further investigation, as one might naively rather expect a
monotonic change with doping.

\subsection{Temperatures below~$T_c$}

\begin{figure}
\centerline{\epsfxsize=8cm\epsfbox{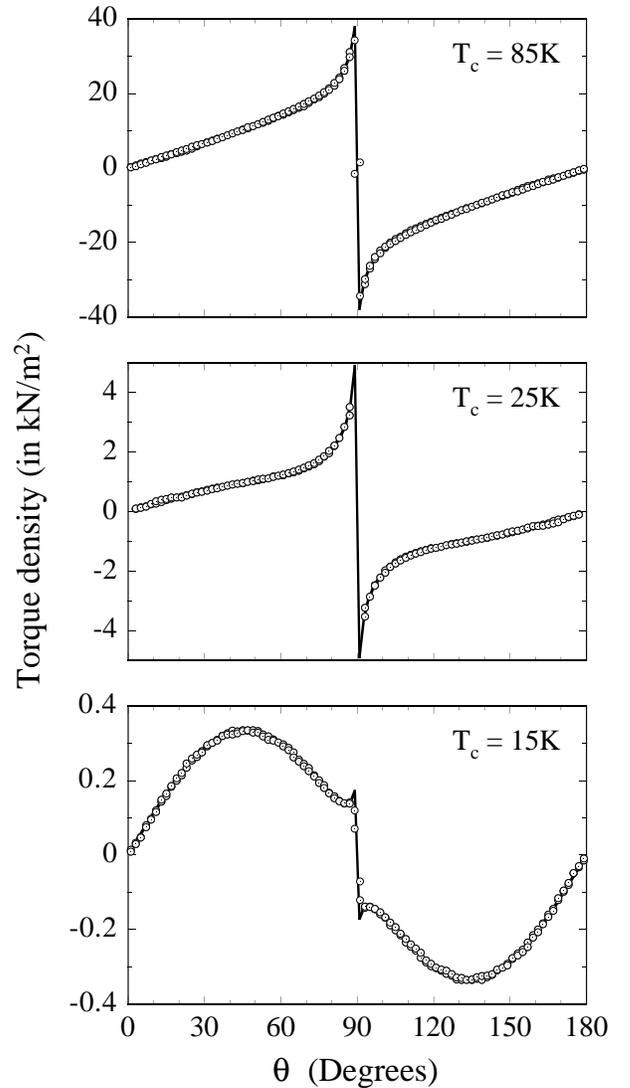}}
\caption{Typical angular dependence of the torque density for the
  crystals investigated in this work. In all cases, \mbox{$T/T_c
    \simeq 0.85$} and \mbox{$H = 5$T}. The solid curves are the fits
  according to Eq.~(\ref{a1a2a3}).}
\label{typical}
\end{figure}

Torque density data for all three crystals, at \mbox{$T/T_c = 0.85$}
and \mbox{$H = 5$T}, are shown in Fig.~\ref{typical}. At a reduced
temperature of 0.85, a field of 5T along the $c$-axis is much greater
than $H_{\rm irr}$ for all three crystals (see Fig.~\ref{irrfields}).
Irreversibility is therefore confined to angles very close
to~\degrees{90}.

There is a sharp quantitative distinction between the three cases. The
torque densities for the \mbox{$T_c = 15$K}, 25K, and 85K crystals
stand in the approximate ratios 1\,:\,10\,:\,100. The angular
dependences for the \mbox{$T_c = 25$K} and 85K crystals are in at
least qualitative accord with Eq.~(\ref{kogantorque}) (see
Fig.~\ref{koganplot}). However, the angular dependence for the
\mbox{$T_c = 15$K} crystal is quite different, containing a strong
contribution that appears to vary as \mbox{$\sim\sin 2\theta$}.

We therefore tried fitting all the data to an expression of the form
\begin{equation} \label{a1a2a3}
  \tau(\theta) = A_1 \sin 2\theta  +  A_2 \sin\theta  +  A_3
  \sin\theta\,\ln|\!\cos\theta|
\end{equation}
for \mbox{$0 < \theta < \pi/2$}. For $\theta$ lying in the second
quadrant, the torque is given by \mbox{$\tau(\theta) = -
  \tau(\pi-\theta)$}, in conformity with its antisymmetry around the
$ab$-plane. The sum of the $A_2$- and $A_3$-terms is equivalent to
\mbox{$- A_3 \sin\theta\,\ln\left( \eta H_{c2} / H|\!\cos\theta|
  \right)$}, with an apparent critical field \mbox{$\eta H_{c2} = H
  \exp(-A_2/A_3)$}, and therefore corresponds to the pure vortex
torque expected from Eq.~(\ref{kogantorque}). The linear decomposition
in Eq.~(\ref{a1a2a3}) provides the most convenient representation of
our data in that it facilitates error estimates and in that the fitting
coefficients have the same units (N/m$^2$) as the torque density
itself. $A_1$, $A_2$, and $A_3$ were adjusted to give a least squares
fit to the data at each field and temperature. A central result of our
work is that Eq.~(\ref{a1a2a3}) provides an excellent empirical fit to
all the data (see, e.g., Fig.~\ref{typical}). The fit parameters for
all temperatures, fields, and samples examined in this study have been
tabulated in Table~\ref{fitcoeff} and therefore provide a
comprehensive representation of our experimental results. In all
cases, the fit is within experimental uncertainties in the region
\mbox{$0 < \theta < \theta_{\rm irr}$}, where the irreversibility
angle $\theta_{\rm irr}$ can be found using Eq.~(\ref{thetairr}) and
the $H_{\rm irr}$ data shown in Fig.~\ref{irrfields}.

\subsection{Temperatures above~$T_c$} \label{normalstate}

For all three crystals, both the coefficients $A_2$ and $A_3$ in
Table~\ref{fitcoeff}, corresponding to the superconducting vortex
torque, extrapolate to zero at $T_c$, but this is not the case for the
coefficient $A_1$: for the \mbox{$T_c = 15$K} and 25K crystals, the
\mbox{$\sin 2\theta$} term could be extracted for temperatures well
above $T_c$.  However, for the \mbox{$T_c = 85$K} crystal, the
contribution of the \mbox{$\sin 2\theta$} term is very small compared
to the magnetoresistive background (see Appendix~\ref{errest}), and no
systematic data could be obtained above~$T_c$.

The sinusoidal torque \mbox{$A_1\sin 2\theta$} in Eq.~(\ref{a1a2a3})
corresponds to a magnetization~$M$ that is proportional to the
effective field \mbox{$H\cos\theta$} along the $c$-axis, with a linear
susceptibility of \mbox{$\chi = -2\mu_0A_1/H^2$}. (More precisely, if
the additional effects of small normal state paramagnetism are taken
into account, this quantity is the susceptibility {\em anisotropy\/}
\mbox{$\Delta\chi = \chi_c - \chi_{ab}$}.) For Eq.~(\ref{a1a2a3}) to
consistently describe the torque data at different applied fields, the
coefficient~$A_1$ therefore has to be proportional to~$H^2$.

\begin{figure}
\centerline{\epsfxsize=8cm\epsfbox{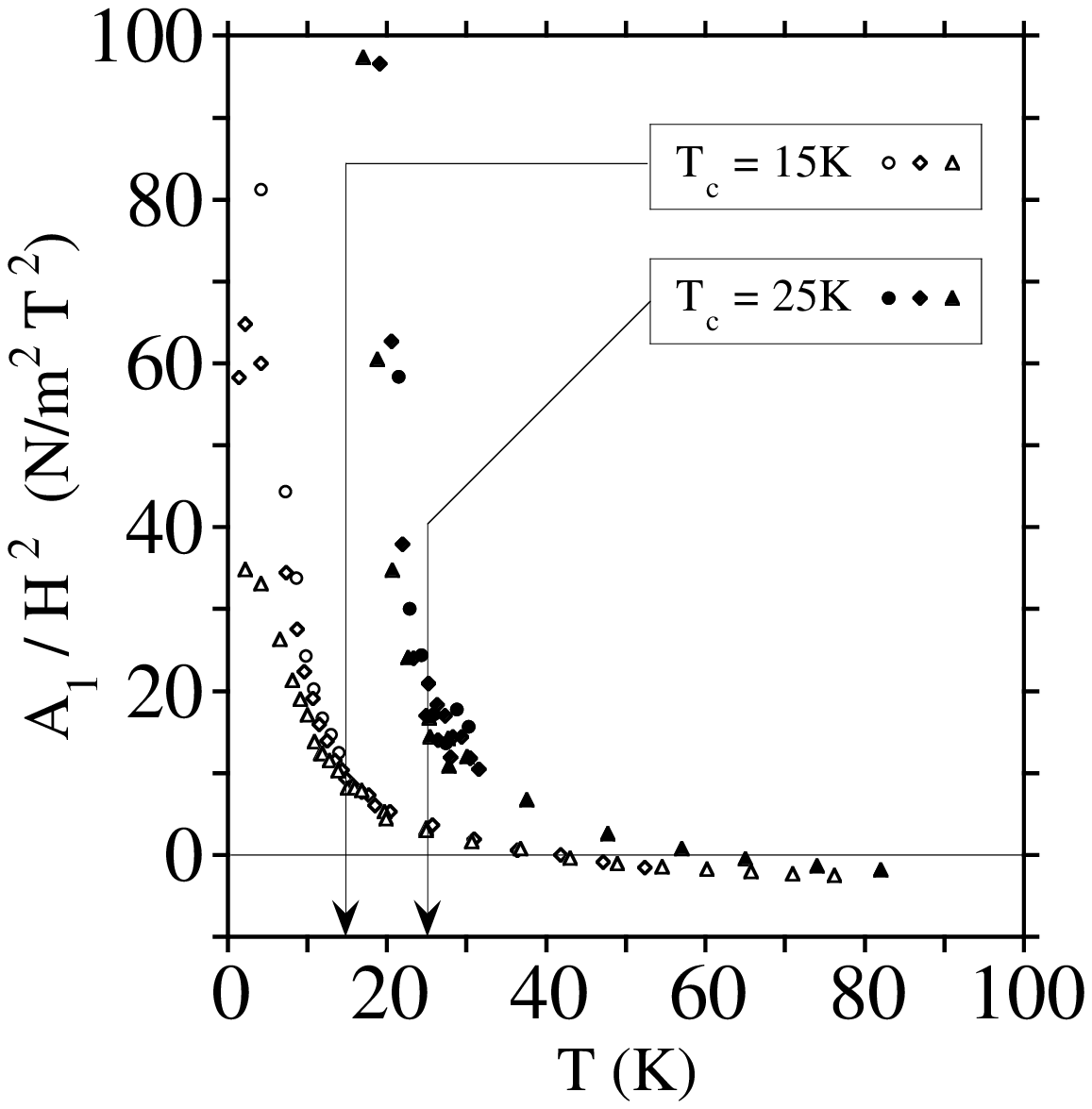}}
\caption{Temperature dependence of the normalized \mbox{$\sin
    2\theta$} coefficient \mbox{$A_1/H^2 = -\Delta\chi/2\mu_0$}, for
  the \mbox{$T_c = 15$K} and 25K crystals, at \mbox{$H = 3$T}
  (circles), 5T (diamonds), and 10T (triangles).  (The \mbox{$H = 1$T}
  data has been omitted as it is affected by the onset of
  irreversibility close to the $ab$-plane below~$T_c$.) The
  magnetoresistive error is estimated to be less than
  0.5\,N/m$^2$T$^2$ at all temperatures, cf.\ the Appendix.}
\label{slopedata}
\end{figure}

Fig.~\ref{slopedata} shows data for the temperature dependence of
\mbox{$A_1 / H^2$} for both the \mbox{$T_c = 15$K} and 25K crystals at
different fields and is indeed in basic agreement with \mbox{$A_1\sim
  H^2$}.  Deviations from an $H^2$-dependence evidently occur at
temperatures below $T_c$, probably due to slight systematic variations
between the runs at different applied fields, with a comparatively
large effect on the outcome of the non-orthogonal three-parameter fit
in Eq.~(\ref{a1a2a3}).

\section{Discussion}

Data for the angular dependence of the torque in \tlbacod\ over a wide
range of temperature, field, and doping are parametrized in
Table~\ref{fitcoeff}. We now discuss the magnetization $M$, and its
dependence on the field component along the $c$-axis, \mbox{$H_\bot=
  H\cos\theta$}. In other high-$T_c$ superconductors such as
Bi$_2$Sr$_2$CaCu$_2$O$_8$, the London relation \mbox{$M \sim \ln (\eta
  H_{c2} / H_\bot)$} provides a good approximation for the field
dependence of the magnetization.\cite{martinez} Significant deviations
from London behaviour have been noted close to $T_c$, and interpreted
in terms of classical (vortex) fluctuations.\cite{bulaevskii1,kogan2}
Smaller deviations at lower temperatures have been interpreted in
terms of quantum fluctuations.\cite{martinez,bulaevskii2} As we now
discuss, the reversible magnetization of overdoped \tlbacod, however,
departs drastically from London behaviour for all temperatures and
fields.

\begin{figure}
\centerline{\epsfxsize=8.7cm\epsfbox{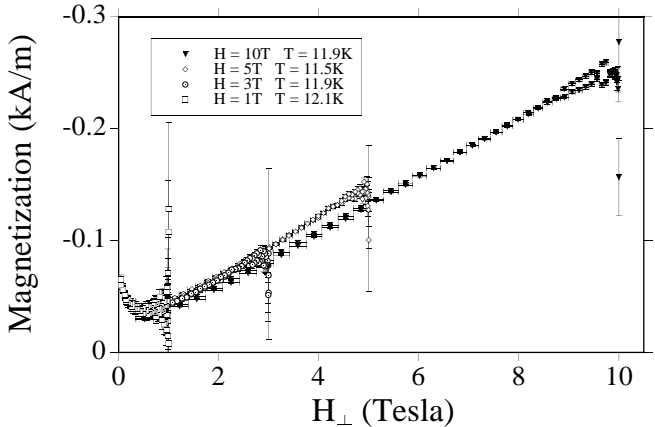}}
\caption{Diamagnetic magnetization plotted against
  the applied field component $H_\bot$ along the $c$-axis, for the
  \mbox{$T_c = 15$K} sample at \mbox{$T\simeq 12$K}. The data were
  extracted from the $\tau(\theta)$ measurements at applied fields~$H$
  of 1T, 3T, 5T, and~10T. For \itH\ close to the $c$-axis, i.e.\ for
  \mbox{$H \simeq H_\bot$}, the measurement uncertainty associated
  with a single data point diverges as \mbox{$1/\sin\theta$}.}
\label{fireworks}
\end{figure}

The magnetization $M(H_\bot)$ can be retrieved from the $\tau(\theta)$
measurements, using the relation \mbox{$\tau = MH\sin\theta$}. This
assumes that $M$ lies close to the $c$-axis, which certainly holds for
vortex magnetization (but might be invalidated by strong in-plane
normal state paramagnetism). As an example, Fig.~\ref{fireworks} shows
the magnetization data on the \mbox{$T_c = 15$K} crystal at a
temperature of \mbox{$T\simeq 12$K}, as retrieved from the
$\tau(\theta)$ measurements at applied fields~$H$ of 1T, 3T, 5T,
and~10T. 

Unfortunately, the temperatures in Fig.~\ref{fireworks} do not
coincide perfectly; moreover, data points with \itH\ pointing
\begin{table}
\addtolength{\textheight}{0.2cm}
\onecolumn
\addtolength{\topmargin}{-0.2cm}
\caption{Compilation of the fit parameters to the experimental
  $\tau(\theta)$ curves, according to Eq.~(\ref{a1a2a3}), for all
  three samples and all fields and temperatures employed. This table
  provides a comprehensive representation of our experimental results.
  The fit was performed over the whole reversible region of the
  experimental data (except for a region of \mbox{$\pm$\degrees{3}}
  around the $ab$-plane where $\tau(\theta)$ was too steep).
  The random and systematic errors in the parameters are comparable
  to those of individual data points, discussed in the
  Appendix~\ref{errest}, but the systematic
  error is enhanced by a factor of \mbox{$\sim 7$} for $A_1$ and
  $A_3$, and \mbox{$\sim 15$} for $A_2$, due to the nonorthogonality
  of the fit functions employed in Eq.~(\ref{a1a2a3}). For partly
  irreversible $\tau(\theta)$, the random errors get worse because of the
  restricted fitting range.}
\label{fitcoeff}
\begin{tabular}[h]{cc|ccc||cc|ccc}
$T_c = 15$K & \multicolumn{1}{c}{$T$ (K)} & $A_1$ (kN/m$^2$) & $A_2$ (kN/m$^2$) & $A_3$ (kN/m$^2$) &
$T_c = 25$K & \multicolumn{1}{c}{$T$ (K)} & $A_1$ (kN/m$^2$) & $A_2$ (kN/m$^2$) & $A_3$ (kN/m$^2$) \\
\hline
$B = 1$T &  4.2 & -0.069 &  0.356 & -0.028 & $B = 1$T & 16.5 & 0.32 &  1.19 & -1.58 \\
         &  7.3 & -0.004 &  0.111 & -0.081 &          & 19.8 & 0.17 &  0.51 & -0.76 \\
         &  8.8 &  0.031 &  0.024 & -0.084 &          & 22.9 & 0.09 &  0.17 & -0.22 \\
\cline{6-10}               
         &  9.9 &  0.021 &  0.021 & -0.053 & $B = 3$T & 6.6  & 5.62 & -4.22 & -15.6 \\
         & 11.0 &  0.009 &  0.025 & -0.029 &          & 8.3  & 4.49 & -2.93 & -13.3 \\
         & 12.1 &  0.015 &  0.009 & -0.017 &          & 9.8  & 3.67 & -2.10 & -11.4 \\
         & 13.2 &  0.013 &  0.003 & -0.008 &          & 11.2 & 2.90 & -1.36 & -9.59 \\
         & 14.3 &  0.011 &  0.001 & -0.004 &          & 12.6 & 2.44 & -1.13 & -8.15 \\
\cline{1-5}                                 
$B = 3$T &  4.2 &  0.731 & -0.590 & -0.724 &          & 14.1 & 2.00 & -0.95 & -6.62 \\
         &  7.3 &  0.400 & -0.312 & -0.376 &          & 15.6 & 1.93 & -1.35 & -5.57 \\
         &  8.7 &  0.304 & -0.208 & -0.251 &          & 17.0 & 1.69 & -1.35 & -4.38 \\
         &  9.8 &  0.219 & -0.093 & -0.146 &          & 18.5 & 1.14 & -0.71 & -2.99 \\
         & 10.8 &  0.182 & -0.060 & -0.094 &          & 20.0 & 0.93 & -0.62 & -2.04 \\
         & 11.9 &  0.150 & -0.035 & -0.053 &          & 21.5 & 0.53 & -0.10 & -1.09 \\
         & 13.0 &  0.132 & -0.024 & -0.029 &          & 22.9 & 0.27 &  0.20 & -0.48 \\
         & 14.0 &  0.112 & -0.008 & -0.012 &          & 24.4 & 0.22 &  0.16 & -0.23 \\
\hline
$B = 5$T &  1.4 &  1.46  & -0.04  & -0.31  & $B = 5$T &  4.2 & 13.0 & -14.7 & -26.5 \\
         &  2.3 &  1.62  & -0.54  & -0.67  &          &  7.0 &  8.6 &  -9.2 & -19.5 \\
         &  4.2 &  1.50  & -1.00  & -0.93  &          &  7.7 &  8.4 &  -9.2 & -18.9 \\
         &  7.4 &  0.86  & -0.54  & -0.48  &          &  9.3 &  7.9 &  -9.5 & -17.6 \\
         &  8.8 &  0.69  & -0.37  & -0.33  &          & 10.7 &  7.3 &  -9.2 & -15.9 \\
         &  9.7 &  0.56  & -0.22  & -0.22  &          & 12.1 &  6.5 &  -8.3 & -13.8 \\
         & 10.7 &  0.48  & -0.15  & -0.15  &          & 13.5 &  6.1 &  -8.1 & -12.2 \\
         & 11.5 &  0.40  & -0.07  & -0.08  &          & 14.9 &  5.8 &  -7.9 & -10.5 \\
         & 12.5 &  0.35  & -0.04  & -0.05  &          & 16.3 &  4.1 &  -5.2 &  -7.7 \\
         & 13.5 &  0.29  &  0.01  & -0.01  &          & 17.7 &  3.2 &  -3.9 &  -5.7 \\
         & 14.4 &  0.26  &  0.02  &  0.00  &          & 19.1 &  2.4 &  -2.7 &  -4.0 \\
\cline{1-5}                                 
$B = 10$T&  1.4 &  3.64  & -1.52  & -1.24  &          & 20.6 &  1.6 &  -1.4 &  -2.4 \\
         &  2.2 &  3.48  & -1.24  & -1.02  &          & 22.0 &  1.0 &  -0.4 &  -1.3 \\
         &  4.2 &  3.32  & -1.31  & -1.02  &          & 23.4 &  0.6 &   0.1 &  -0.6 \\
         &  6.6 &  2.63  & -0.94  & -0.68  &          & 24.9 &  0.4 &   0.2 &  -0.2 \\
\cline{6-10}
         &  8.2 &  2.13  & -0.61  & -0.44  & $B = 10$T&  2.5 & 16.3 & -13.4 & -23.6 \\
         &  9.2 &  1.90  & -0.44  & -0.32  &          &  9.9 & 21.1 & -29.8 & -28.2 \\
         & 10.0 &  1.71  & -0.32  & -0.23  &          & 11.7 & 18.6 & -26.5 & -24.0 \\
         & 10.9 &  1.39  & -0.06  & -0.10  &          & 13.5 & 15.1 & -21.0 & -18.9 \\
         & 11.7 &  1.24  &  0.03  & -0.04  &          & 15.2 & 12.7 & -17.2 & -15.0 \\
         & 11.9 &  1.24  &  0.02  & -0.04  &          & 17.0 &  9.7 & -12.5 & -10.8 \\
         & 12.8 &  1.15  &  0.02  & -0.02  &          & 18.8 &  6.1 &  -6.4 &  -6.2 \\
         & 13.9 &  1.03  &  0.02  & -0.00  &          & 20.7 &  3.5 &  -2.3 &  -3.0 \\
         &      &        &        &        &          & 22.7 &  2.4 &  -0.8 &  -1.4 \\
         &      &        &        &        &          & 25.4 &  1.5 &   0.4 &  -0.2 \\
\hline\hline
$T_c = 85$K & \multicolumn{4}{c}{}                  & $T_c = 85$K & \multicolumn{4}{c}{} \\
\hline
$B = 1$T & 35.3 &  -2.0  &  16.7  & -1.7   & $B = 5$T & 50.7 &  4.8 &  20.2 & -17.2 \\
         & 44.4 &  -0.8  &  12.5  & -2.9   &          & 60.5 &  3.3 &  15.2 & -12.3 \\
         & 53.4 &  -0.4  &   9.5  & -2.8   &          & 69.8 &  1.2 &  10.6 &  -6.5 \\
         & 62.4 &   0.1  &   6.2  & -2.4   &          & 79.0 & -0.7 &   6.5 &  -0.7 \\
\cline{6-10}
         & 71.4 &   1.1  &   1.8  & -1.8   & $B = 10$T& 54.8 &  6.8 &  27.8 & -27.7 \\
\cline{1-5}                                           
$B = 3$T & 44.5 &   2.9  & 18.9   &-12.2   &          & 65.4 &  2.6 &  20.9 & -16.6 \\
         & 53.7 &   2.0  & 15.4   & -9.4   &          & 70.8 & -0.1 &  18.1 & -10.2 \\
         & 63.3 &   1.4  & 10.7   & -6.5   &          & 76.0 & -0.6 &  12.9 &  -5.1 \\
         & 72.4 &   0.4  & 6.47   & -3.2   &          &      &      &       &
\end{tabular}
\addtolength{\textheight}{-0.2cm}
\twocolumn
\addtolength{\topmargin}{0.2cm}
\end{table}
\noindent close to the $c$-axis (i.e.\ for \mbox{$H \simeq H_\bot$}) have error
bars diverging as \mbox{$1/\sin\theta$}. Therefore, to get a more
systematic and comprehensive representation of our data, we employed
an averaging technique. At a given temperature, instead of working
with individual data points, we retrieved the $\tau(\theta)$ curves
from Eq.~(\ref{a1a2a3}), where the fit coefficients $A_1$, $A_2$,
and~$A_3$ were interpolated at the desired temperature from
Table~\ref{fitcoeff}. The resulting four $M(H_\bot)$ curves at the
applied fields~$H$ of 1T, 3T, 5T, and~10T were then averaged. (In
order to minimize the overall uncertainty, a weighting factor of
$1/\sigma^2(H, H_\bot)$ was assigned to each curve, where $\sigma(H,
H_\bot)$ is the estimated uncertainty of the extracted magnetization
-- which diverges as \mbox{$H_\bot \to H$}).

\begin{figure}
\onecolumn
\centerline{\epsfxsize=18cm\epsfbox{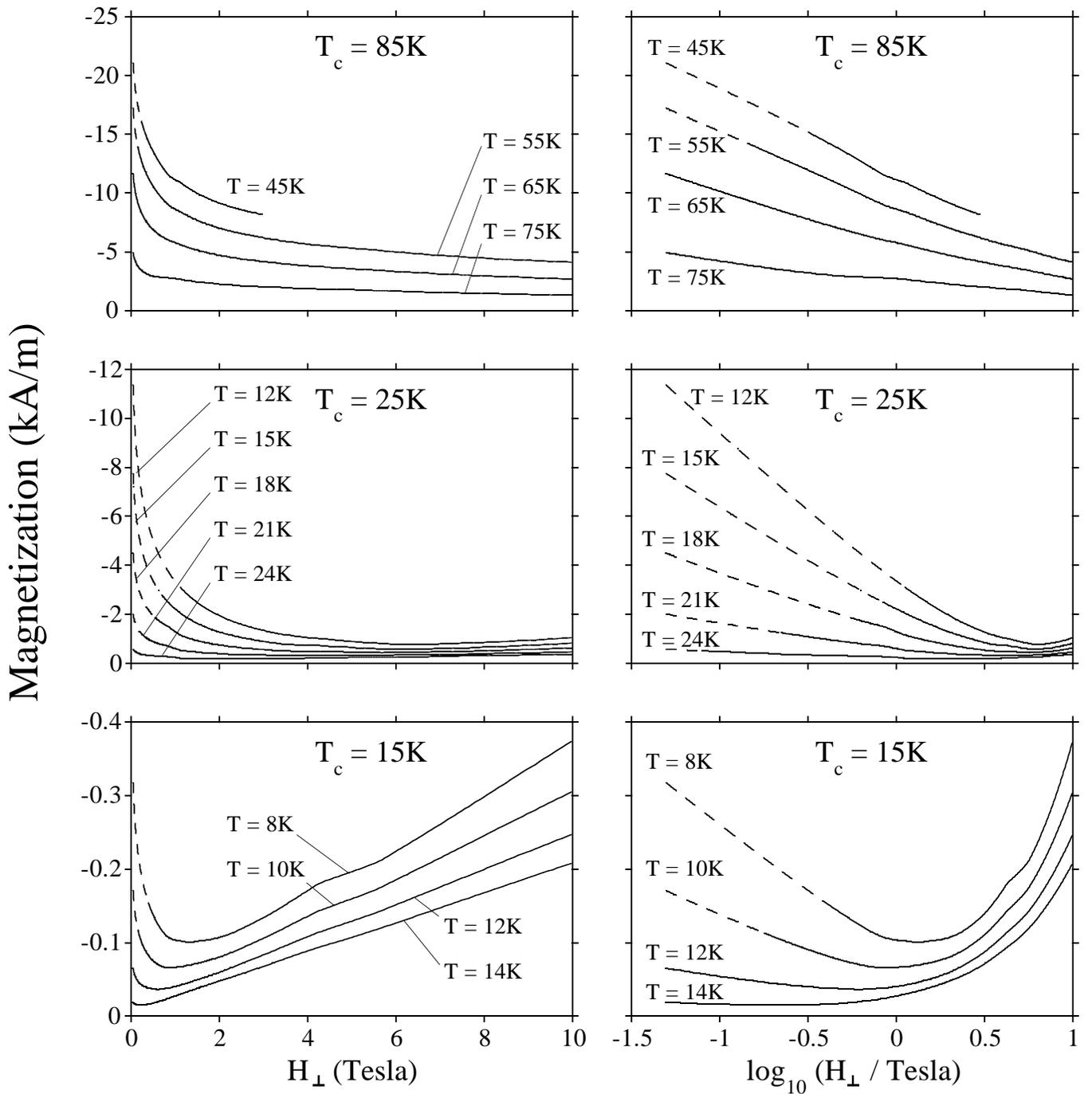}}
\caption{Reversible diamagnetic magnetization (solid lines) and its
  extrapolation into the irreversible region (dashed), plotted against
  the applied field component $H_\bot$ along the $c$-axis (left
  panels) and its logarithm (right panels), at different temperatures
  for all three samples. The data were interpolated from the fit
  parameters in Table~\ref{fitcoeff}, assuming that the magnetization
  is always parallel to the $c$-axis. The estimated error is around
  \mbox{$\pm 2\,$A/m} for the \mbox{$T_c = 15$K} sample, \mbox{$\pm
    15\,$A/m} for \mbox{$T_c = 25$K}, and \mbox{$\pm 50\,$A/m} for
  \mbox{$T_c = 85$K}. The ``wiggles'' in the lines are artefacts,
  caused by slightly different values of~$M$ obtained from runs in the
  experimentally applied fields of 1T, 3T, 5T, and~10T.}
\label{MvsH}
\twocolumn
\end{figure}

This averaged magnetization is plotted against $H_\bot$ and
\mbox{$\log H_\bot$} in Fig.~\ref{MvsH} for a variety of temperatures
below~$T_c$ for all three crystals. The dashed lines in all the panels
represent estimates for the equilibrium magnetization in fields {\em
  below\/} the irreversibility field. These estimates were obtained by
extrapolating the fit in Eq.~(\ref{a1a2a3}) to angles higher
than~$\theta_{\rm irr}$ and therefore to applied fields \mbox{$H_\bot
  < H_{\rm irr}$}. The reader should ignore the small ``wiggles'' in
the lines: these are artefacts, caused by slightly different values
of~$M$ obtained from runs in the experimentally applied fields of 1T,
3T, 5T, and~10T.

Consider the right panels in Fig.~\ref{MvsH}, showing the variation of
the magnetization with \mbox{$\log H_\bot$}. For the \mbox{$T_c =
  85$K} crystal, there are some departures from linearity, but the
form of the data is qualitatively similar to that seen in other
cuprates such as Bi$_2$Sr$_2$CaCu$_2$O$_8$.\cite{martinez} For the
\mbox{$T_c = 15$K} and 25K crystals, however, $M(\log H_\bot)$ departs
strongly from linearity and displays an upturn which -- at least at
lower temperatures -- occurs in the vicinity of the irreversibility
field. As can be seen in the left panels of Fig.~\ref{MvsH}, there
then exists an extensive field range over which \mbox{$M \sim H_\bot$}
(corresponding to contributions from the first term in
Eq.~(\ref{a1a2a3})). This region extends as the temperature rises, and
above~$T_c$, the magnetization is linear over the whole field range,
as discussed in section~\ref{normalstate}.

A (roughly) linear magnetization, \mbox{$M \sim H_\bot$}, has been
observed previously in other high-$T_c$ materials in the immediate
vicinity of $T_c$, and ascribed to classical thermal
fluctuations.\cite{bulaevskii1} Its appearance at very much lower
temperatures is a surprising result and raises the possibility that
some sort of artefact might be affecting the measurements. A linear
relation between $M$ and~$H$ is characteristic of both the Pauli
susceptibility in a normal metal and a linear Curie spin
susceptibility. All high-$T_c$ materials exhibit such a spin
susceptibility which is anisotropic due to crystal field
interactions,\cite{miljak} leading to a torque that varies as
\mbox{$H^2 \sin 2\theta$}. Such a (normal state) torque mimics the
first term in Eq.~(\ref{a1a2a3}), suggesting that the linear
magnetization may have nothing to do with superconductivity and may
simply reflect the presence of a normal-state susceptibility
anisotropy with a strong temperature dependence.

Although possible in principle, there are two reasons why such a
scenario is unlikely: firstly, the sign of the effect is implausible.
Fig.~\ref{slopedata} shows that the \mbox{$\sin 2\theta$} component
changes sign at higher temperatures, for both the \mbox{$T_c = 15$K}
and 25K crystals. At higher temperatures, the magnitude of the
associated susceptibility anisotropy is \mbox{$\sim 10^{-5}$} (SI
units), a number that is typical for the cuprates;\cite{miljak} the
sign of this term is such that the maximum susceptibility lies along
the $c$-axis, as observed in essentially all other high-$T_c$
materials. However, as shown in Fig.~\ref{slopedata}, the sign is
reversed below \mbox{$T \simeq 2.5 T_c$}, for both crystals. It is
very difficult to understand why a normal state effect of this type
should flip sign, and why it should do so at a temperature linked to
the superconducting transition. In addition, at a given temperature,
the normal state susceptibility in \tlbacod\ is known to be {\em
  larger\/} for crystals with lower critical
temperatures.\cite{shimakawa} The linear susceptibility deduced from
our torque measurements is shown in Fig.~\ref{slopedata}; at a given
temperature, it is {\em smaller\/} for the lower $T_c$ crystal.
Although a normal state origin for the linear magnetization cannot be
completely dismissed, the balance of evidence suggests that it is
likely to be a consequence of superconducting order.

It should be mentioned that the diamagnetic upturn in our
susceptibility anisotropy data in Fig.~\ref{slopedata} exhibits
qualitative similarity to the ``diamagnetic Curie-Weiss law'' observed
in the normal state of YBa$_2$Cu$_3$O$_7$\cite{miljak,lee} and, most
notably, in La$_{2-x}$Sr$_x$Cu$_2$O$_4$,\cite{miljakla} the origin of
which remains unknown. Fitting our $\Delta\chi(T)$ data to a
Curie-Weiss term with a constant background, we obtain
\mbox{$\Theta_{\rm 25K} \simeq 15$K} and, rather intriguingly,
\mbox{$\Theta_{\rm 15K} \simeq 0$} as characteristic temperatures for
the two samples, where the subscript refers to the critical temperature.

It is also interesting to note that, independent of the origin of the
\mbox{$\sin 2\theta$} contribution in overdoped \tlbacod, subtracting
it from the data does not lead to a standard superconducting torque.
Although its angular dependence is qualitatively described by
Eq.~(\ref{kogantorque}), 
\noindent attempts to obtain the apparent upper
critical field \mbox{$\eta H_{c2}$} gave unphysical temperature
dependences. This is not the case for other high-$T_c$
superconductors, where physically reasonable values and temperature
dependences of $H_{c2}$ can be extracted from torque
data.\cite{vulcanescu,zech}

If all the data shown in Fig.~\ref{MvsH} are indeed directly linked to
superconducting order, they give a new perspective on one of the
problems that motivated our study, that of the upper critical field in
strongly overdoped \tlbacod. Recently, an attempt has been
made to deduce the temperature dependence of $H_{c2}$ from a Raman
scattering experiment on a \tlbacod\ crystal with \mbox{$T_c \simeq
  25$K}.\cite{blumberg} In our \mbox{$T_c = 25$K} sample, our data
suggest that diamagnetism persists to much higher fields than those
identified as $H_{c2}$ in that study.

Our data also clearly imply the existence of diamagnetism -- and thus
remnant superconducting order -- at fields and temperatures for which
transport measurements\cite{hstar} suggest that the normal state has
been fully restored. This conclusion was previously reached by
Carrington et al.\ on the basis of heat capacity
measurements,\cite{carrington2} but the field dependence of the heat
capacity anomaly in low-$T_c$ \tlbacod\ appeared to be qualitatively
similar to that seen in other cuprates with higher transition
temperatures. As noted in Ref.~\onlinecite{carrington2}, this is
difficult to reconcile with striking differences in the transport
behaviour between optimally doped and overdoped \tlbacod. The new
result from our torque data is that although remnant superconducting
order exists above the irreversibility line for all doping levels,
departures from London behaviour clearly emerge as $T_c$ is reduced
from 85K to~15K\@.

In summary, the magnetic behaviour of low-$T_c$ crystals belonging to
the \tlbacod\ system has been a puzzle for a number of years.
Transport measurements have suggested that a vortex liquid does not
appear and that the irreversibility and upper critical field
boundaries coincide. On the other hand, specific heat measurements
have shown that superconducting order persists well above the
irreversibility boundary. In this work, we have studied the
superconducting magnetization and demonstrated that significant
departures from London liquid behaviour occur for all three samples
studied, i.e.\ across the entire overdoped region. The departures
increase in severity as the doping level is increased. In particular,
for strongly overdoped \mbox{($T_c = 15$K)} material, a London-like
vortex liquid does not appear above the irreversibility boundary. The
remnant superconducting order is characterized by a linear diamagnetic
response, a response that persists well above $T_c$ and also up to the
highest field employed here (10 Tesla). We hope that this novel result
will encourage further effort to develop our understanding of the
intriguing behaviour of overdoped \tlbacod.

\acknowledgements 

We thank A.~Carrington for useful discussions of unpublished heat
capacity data on two of the crystals studied in this work, and
V.~G.~Kogan, A.~J.~Millis, and A.~M.~Tsvelik for valuable discussions.
We are also grateful to N.~E.~Hussey for contributions to the early
part of the project.  The work was funded by the U.K. EPSRC, and one
of us (APM) gratefully acknowledges the support of the Royal Society.
In addition, one of us (DEF) acknowledges support provided by an EPSRC
Visiting Fellowship and NSF support under grant DMR93-07581.

\appendix
\section{Calibration} \label{calibration}
        
With a force constant of 2.5\,N/m, the angular displacement of the
lever for all the reversible torques encountered in this study is
estimated to be significantly less than \degrees{1}. Under these
circumstances, it can be assumed\cite{privcom} that the change in
lever resistance, \mbox{$\Delta R$}, is proportional to the torque
$\tau$ experienced by the sample, or,
\begin{equation} \label{propsens}
  \tau = \alpha \, \Delta R
\end{equation}
where $\alpha$ is a calibration constant. 

Let $\rho$ and $V$ be the density and volume of the attached crystal,
respectively. To calibrate the cantilever, the balanced bridge
configuration described in the text was rotated in zero magnetic field
where the lever only experiences a gravitational torque given by
\begin{equation} \label{gravtorq}
  \tau_g = \rho\, Vg\,l \sin\theta
\end{equation}
where $\theta$ is the angle to the vertical, $l$ is the length of the
lever, and $g$ is the gravitational acceleration of the Earth. The
resistive path is assumed to be close to the region of maximum stress
at the lever base. The amplitude of the associated resistance change,
\mbox{$\Delta R_g$}, was measured and the sensitivity obtained from
\begin{equation} \label{getpropconst}
  \alpha = \frac{\rho\, Vg\,l}{\Delta R_g} \:.
\end{equation}

The value of $\alpha$ obtained in this way was\linebreak \mbox{$8 \times
  10^{-11} {\rm Nm}/\Omega$}. No systematic study of the temperature
dependence of the sensitivity was performed. However, Yuan et
al.\cite{yuan} found the piezolever sensitivity to be constant to a
factor of 2 over the temperature range involved in this work, a result
that was confirmed for one lever used in this study. The calibration
constant in the present work is therefore subject to an uncertainty of
order \mbox{$\pm 50\%$}.

\section{Error Estimate} \label{errest}

The main source of random measurement uncertainty stemmed from
temperature variations. Although these variations were negligible when
the lever was immersed in liquid helium, they were about \mbox{$\pm
  0.03$K} when it was surrounded by exchange gas. These variations
coupled into the system via the temperature dependent lever resistance
$R(T)$. The compensation arrangement could not eliminate this effect, but at
least suppress it to 5\% or less. Since $dR/dT$ itself varied, the
resulting torque uncertainty depended on temperature and rose from to
a minimum of about \mbox{$2 \times 10^{-13}\,$Nm} at 5K to a maximum
of about $10^{-12}\,$Nm at 30K.

The small contribution from the gravitational signal
(Appendix~\ref{calibration}) was measured and subtracted from the
data. Cantilever magnetoresistance was less easy to deal with and
constituted the most serious systematic uncertainty for the
measurements reported here. Initially varying as $B^2$, the
(temperature-dependent) change in lever
resistance with field becomes almost linear above
0.5T and reaches a value as high as \mbox{$\sim
  100\,\Omega$} in a field of~10T\@. Again, the compensation lever
could trim down the effective change to a few Ohms, but this was still
comparable to the total signal for the \mbox{($T_c = 15$K)} crystal.

The strategy for reducing this uncertainty was two-fold. In the first
place, although the magnetoresistance is large, its anisotropy at
temperatures below 40K is relatively small, reducing the apparent
magnetoresistive torque to \mbox{$6 \times 10^{-11}\,$N/m$^2$T} when
the levers were rotated in the field. By confining measurements to the
angular variation of the torque, the magnetoresistance error was
therefore reduced to a manageable level. (This was the reason why the
irreversibility field was obtained from the angular characteristic,
rather than a traditional field sweep.)

It was further reduced by noting that, in our geometrical
configuration, the symmetry of the vortex torque $\tau_{\rm v}$ and the
apparent magnetoresistive torque $\tau_{\rm m}$ around the $ab$-plane were
odd and even, respectively:
\begin{eqnarray} \label{oddeven}
  \tau_{\rm v}(\pi - \theta) & = & - \tau_{\rm v}(\theta) \\
  \tau_{\rm m}(\pi - \theta) & = &   \tau_{\rm m}(\theta) \nonumber
\end{eqnarray}
        
Given an observed angular dependence $\tau_{\rm raw}(\theta)$, over
two angular quadrants, the vortex torque $\tau_{\rm v}(\theta)$ was
therefore extracted according to:
\begin{equation} \label{symmetrize}
  \tau_{\rm v}(\theta) = \frac{1}{2} \left( \tau_{\rm raw}(\theta) -
  \tau_{\rm raw}(\pi - \theta) \right)
\end{equation}

This is the quantity actually plotted in Figs.~\ref{irrplot}
and~\ref{typical}. This procedure cancelled the symmetric
magnetoresistive component. Misalignments of the levers
($\sim$\degrees{1}) led to a residual torque uncertainty of
\mbox{$\sim 10^{-13}\,$Nm/T} (below 40K; \mbox{$\sim 10^{-12}\,$Nm/T}
at higher temperatures). Note that, in a 10T field, this is about
10$^2$~times larger than the intrinsic uncertainty quoted for the
piezolever method.\cite{rossel} The degree of cancellation could
probably be improved, but it was adequate for the present work. In a
field of 10T, the remaining torque uncertainty corresponds to a moment
uncertainty of \mbox{$\sim 10^{-13}\,$Am$^2$}, about 10$^2$~times
better than can be achieved by commercial SQUID methods.

To compare the above numbers with the torque density results, they
have to be divided by the respective crystal volumes \mbox{$V_{\rm
    15K} = 2.7\times 10^{-13}\,{\rm m}^3$}, \mbox{$V_{\rm 25K} =
  1.3\times 10^{-13}\,{\rm m}^3$}, and \mbox{$V_{\rm 85K} = 0.65\times
  10^{-13}\,{\rm m}^3$}, where the subscript refers to the critical
temperature.

\end{document}